\documentclass[twocolumn,showpacs,amsmath,amssymb,pra,a4paper]{revtex4}
\usepackage{graphicx}
\usepackage{dcolumn}
\usepackage{bm}
\usepackage{color}
\usepackage{amsmath}
\newcommand{\re}{\mathop\mathrm{Re}}
\newcommand{\sech}{ {\rm sech} }

\begin{document}

\title{Spectral Sidebands and Multi-Pulse Formation in Passively Mode Locked Lasers}
\author{Rafi Weill, Alexander Bekker, Vladimir Smulakovsky, and Baruch Fischer}

\affiliation{ Dept.ֿ of Electrical Engineering,
Technion-IIT, Haifa 32000, Israel }

\author{Omri Gat}
\affiliation{Racah Institute of Physics, Hebrew University,
Jerusalem 91904, Israel}

\date{\today}

\begin{abstract}
Pulse formation in passively mode locked lasers is often accompanied
with dispersive waves that form of spectral sidebands due to
spatial inhomogoneities in the laser cavity. Here we present
an explicit calculation of the amplitude, frequency, and precise
shape of the sidebands accompanying a soliton-like pulse. We then
extend the study to the \emph{global} steady state of mode locked
laser with a variable number of pulses, and present experimental
results in a mode locked fiber laser that confirm the theory. The
strong correlation between the temporal width of the sidebands and
the measured spacing between the pulses in multipulse operation
suggests that the sidebands have an important role in the
inter-pulse interaction.
\end{abstract}

\pacs{42.55.Ah, 42.65.Sf, 05.70.Fh}

\maketitle


\paragraph{Introduction}

Soliton-like pulses propagating in nonlinear media with periodic
spatial inhomogeneity can be a source of dispersive waves that are concentrated at resonant frequencies, forming spectral sidebands,
sometimes called Kelly sidebands. \cite{kelly1}. Several papers
\cite{Gordon,kelly1,kelly2,mecozzi} studied the formation and
of dispersive waves by solitons and their propagation in optical fibers with
periodically spaced amplifiers. The same mechanism leads to the
formation of sidebands in mode locked lasers, where inhomogeneities
in the cavity act periodically on the pulse.

When the power of a single pulse saturates the absorber, the steady
state of a passively mode locked laser tends to bifurcate into
configurations where two or more pulses run in the laser cavity
simultaneously. Since the early experiments that demonstrated
multipulse mode locking \cite{mp1,mp2,mp3} it has been observed that
the pulses display a very rich
dynamics, often forming bunches, as a consequence of
complex inter-pulse interactions. The interest in the dispersive
waves in mode locked lasers, beyond their prominent effect on the
pulse shape, arises because they have often been suggested as a
means of inter-pulse interaction in multipulse mode locked lasers
\cite{Grudinin_Gray,Soto-Crespo,Tang,komarov}. Here we focus on the
role of the sidebands in multipulse mode locking.

The absorber saturation leading to multipulse steady states is often
modeled by adding a quintic term to the equations of motion. In former papers \cite{stepsPRL,stepsPRE} we applied the
statistical light-mode dynamics (SLD) theory to this system, and
showed that multipulse mode locking is in effect a series of first
order phase transitions. SLD uses the methods of statistical physics
to analyze the dynamics of the interacting many body light mode
system at an effective finite temperature generated by cavity noise
\cite{GFPRL,GFOC,GGFPRE}. Here we apply the SLD gain balance method
\cite{PRL-2006} to derive the multipulse steady states
with dispersive waves of mode locked lasers with cavity
inhomogeneities.

Our theoretical analysis is based on the master equation of
mode-locked soliton lasers \cite{HausReview,KutzReview}, with an
additive noise term \cite{haus-mecozzi}, where the inhomogeneities
in the cyclic light propagation in the cavity are modeled by a
periodic modulation of the gain and the saturable absorption. We
first study the sidebands in a single-pulse steady state and show
that, unlike free fiber sidebands, the mode locked laser sidebands
reach a steady state with a well-defined bandwidth and a Lorentzian
shape; in real time the dispersive waves form a wide pedestal with
exponentially decaying tails. In particular we demonstrate how the
overall phase of each sideband depends on the relative phase of the
gain and loss modulations. The theory is firmly supported by experimental
observations in mode locked fiber lasers.

Next we derive a nonlinear equation for the global steady state of
the laser, that includes a number of pulses and their accompanying
pedestals, by applying the gain balance principle to the
pulses, sidebands, and cw components of the waveform. We find that the sideband
intensity and the pedestal width increase by a large factor when the
pumping is increased with a constant number of pulses, and then
decrease abruptly when another pulse is formed, so that the
properties of single-pulse sidebands display an oscillatory,
approximately periodic, dependence on the pump power. These
theoretical results are again favorably compared with experiments in
a mode locked fiber laser. We conclude by considering the
implication of our results on the nature of sidebands-mediated
interactions.

\paragraph{Theoretical model}
Our theoretical analysis is based on the mode-locking master
equation model for soliton lasers \cite{HausReview,KutzReview,
haus75}, where the dominant dynamical processes are chromatic
dispersion and Kerr nonlinearity, whose coefficients can be
nondimensionalized by an appropriate choice of units of time and
power. The master equation then takes the form
\begin{align}\label{eq:master}
i\partial_z\psi&=-\partial_t^2\psi-2|\psi|^2\psi\nonumber\\&
+ip_g(z)g[\psi]\left(\psi+\gamma\partial_t^2
\psi\right)\nonumber\\&+ ip_s(z)(s(|\psi|^2)-l)\psi+\Gamma(z,t)
\end{align}
The first two terms on the right-hand-side are the aforementioned
dispersion and Kerr nonlinearity in soliton units. The next term
models the saturable gain of the laser amplifier. $g[\psi]$ is the
overall gain coefficient, and the square brackets signify that $g$
depends on the entire waveform $\psi(t)$ rather than the
instantaneous value of $\psi$ only, and $\gamma$ is the coefficient
of parabolic spectral filtering. We will assume that the the gain
saturation is slow compared to the cavity round trip time $t_R$, so
that $g$ is determined by the overall power
$P=\int\limits_{-t_R/2}^{t_R/2}|\psi|^2 dt$ in the usual manner
\begin{equation}\label{eq:gain}
g[\psi]= \frac{g_0}{1+P/P_s}
\end{equation}
where $g_0$ and $P_s$ are the small signal gain and the saturation
power, respectively.

The next term on the right-hand-side of the master equation models
the fast saturable absorber with transmissivity $s(|\psi|)-l$, where
$l$ is the small signal loss, and $s(0)=0$ by definition. We do not
assume a particular form for the transmissivity function other than
that $s(0)<l$ and that it increases linearly at zero, $s'(0)>0$. We assume for
simplicity that the saturable absorber accounts for all the losses
in the cavity. The final term is a Gaussian white noise source with
covariance
\begin{equation}\label{eq:gamma}
\langle \Gamma^*(z,t)\Gamma(z',t')\rangle=2
T\delta(z-z')\delta(t-t'),
\end{equation}
where the constant $T$ is the rate of internal and injected noise
power. As mentioned above, the noise is a significant factor in the
determination of the steady state; we conjecture that it is also an
essential ingredient in the inter-pulse interaction. The sidebands
are formed by the spatial inhomogeneity of the gain and loss
processes in the laser, described the by $p_g(z)$ and $p_s(z)$
(respectively) that are periodic functions of the cavity roundtrip
length $L$, normalized to $\frac1L\int_0^Ldzp_{g,s}dz=1$.

The dominance of the dispersive effects means that the gain and loss
terms in Eq.\ (\ref{eq:master}) are proportional to a small
parameter, and that the noise term is proportional to an independent
small parameter. In spite of their smallness, the gain, loss and
noise are the crucial terms for the mode locking phenomena discussed
here. At the same time, they also perturb the pulse properties that
are dominated by the dispersive terms; these small perturbation will
be neglected.

\paragraph{Single-pulse side bands}
We begin our analysis assuming conditions under which there is a
single pulse in the cavity with fixed parameters in the steady
state. Since the dominant terms in the master equations are the
dispersion and Kerr effect, the pulse waveform is approximately that
of a nonlinear Schr\"odinger (NLS) soliton. Solitons are defined by
four parameters, amplitude $a$, frequency, timing and phase. The
gain and loss terms in the master equation fix the frequency to
zero, and we can set the timing and phase to zero by an appropriate
choice of origin, so that the soliton waveform is
\begin{equation}\label{eq:sech}
\psi_p(t,z)=a\textrm{sech}(at)e^{i a^2z}
\end{equation}

The soliton waveform $\psi_s$ is not an exact solution of the master
equation because of the gain, loss and noise terms. We therefore
look for a solution of the form
\begin{equation}
\psi(t,z)=\psi_p(t,z)+\psi_b(t,z)+\psi_c(t,z)
\end{equation}
that in addition to the pulse waveform, consists of the sidebands
waveform $\psi_b$ geneated by the cavity inhomogeneities, and the
continuum wave $\psi_c$ generated by the cavity noise. The three
waveform components have different characteristic time scales: the
sidebands are narrow resonances whose temporal width is, as shown
below, inversely proportional to the gain-loss small parameter. This
width is large compared with the pulse width, but small compared
with the cavity roundtrip time---the scale of the continuum.

Both $\psi_b$ and $\psi_c$ have low peak power, and can therefore be
analyzed by the linearized master equation, although their total
power $\int dt|\psi_b|^2$,  $\int
dt|\psi_b|^2$ can be of the order $P$. The effect of noise is
negligible on $\psi_b$, that therefore satisfies the equation
\begin{align}\label{eq:pert1}
\partial_z\psi_b&=i(\partial_t^2\psi_b+4|\psi_p|^2\psi_b+2\psi_p^2\psi_b^*)
\\&
+p_g(z)g[\psi]\left(1+\gamma\partial_t^2\right)(\psi_p+\psi_b)
\nonumber\\&+ p_p(z)(s(|\psi_p|^2)-l)(\psi_p+\psi_b)\nonumber
\end{align}
The right-hand-side of Eq.\ (\ref{eq:pert1}) retains terms that are
of higher order of smallness than $\psi_b$; these terms are in
effect not negligible for $|t|\gg a$ where $\psi_p$ is itself small,
and play a crucial role in the shaping of the side bands, as shown
below.

The discrete modes of the real-linear Eq.\ (\ref{eq:pert1}) express
small variations of the pulse parameters \cite{haus-mecozzi,kfg10},
while $\psi_b$, that consists of radiation emitted by the pulse, is
a linear combination
\begin{equation}\label{eq:alpha}
\psi_b(t,z)=e^{ia^2z}\int\frac{d\omega}{2\pi}(\alpha_\omega(z)u_\omega(t)+\alpha_\omega(z)^*v_\omega(t)^*)
\end{equation}
of the first component of scattering states of the linear operator $L$,
\begin{align}\label{eq:sceigen}
L\begin{pmatrix}u_\omega\\v_\omega\end{pmatrix}&=(-(i+g\gamma)\omega^2-ia^2))\begin{pmatrix}u_\omega\\v_\omega\end{pmatrix}\\
L\begin{pmatrix}v_\omega^*\\u_\omega^*\end{pmatrix}&=((i-g\gamma)\omega^2+ia^2)\begin{pmatrix}v_\omega^*\\u_\omega^*\end{pmatrix}
\end{align}
that acts on two-component wave functions as
\begin{equation}
L=\begin{pmatrix}A&B\\B^*&A^*\end{pmatrix}
\end{equation}
with
\begin{equation*}
A=(i+g\gamma)\partial_t^2-ia^2+4ia^2\sech^2(at)+s(a^2\sech^2(at))
\end{equation*}
and $B=2ia\sech^2(at)$.
It will be argued below that we can use approximate the
space-dependent linear operator acting on $\psi_b$ in Eq.\
(\ref{eq:pert1}) by its space average $L$. Within this approximation
the coefficients defined in Eq.\ (\ref{eq:alpha}) evolve according
to
\begin{align} \nonumber
\partial_z\alpha_\omega=&(-(i+g\gamma)\omega^2-ia^2)\alpha_\omega+(g\alpha_\omega
+\beta_{g\omega}) p_g(z)\\&+(-l \alpha_\omega+\beta_{s\omega})p_s(z)\label{eq:dza}
\end{align}
where $\beta_{g}$ and $\beta_{s}$ are the expansion coefficients of
the forcing terms $g(1+\gamma\partial_t^2)(a\sech(at))$ and
$\big(s(a^2\sech^2(at))-l\bigr)a\sech(at)$ (respectively).
As usual, the expansion coefficients are extracted by inner product
with the adjoint eigenfunctions defined by
\begin{align}
L^\dagger\begin{pmatrix}\bar u_\omega\\\bar v_\omega\end{pmatrix}&=((i-g\gamma)\omega^2+ia^2)(\omega^2+a^2)\begin{pmatrix}\bar u_\omega\\\bar v_\omega\end{pmatrix}\\
L^\dagger\begin{pmatrix}\bar v_\omega^*\\\bar
u_\omega^*\end{pmatrix}&=(-(i+g\gamma)\omega^2-ia^2))\begin{pmatrix}\bar
v_\omega^*\\\bar u_\omega^*\end{pmatrix}
\end{align}
normalized so that
\begin{align}
&\int dt(\bar u_\omega(t)^*u_{\omega'}(t)+\bar v_{\omega}(t)^*v_{\omega'}(t))=2\pi\delta(\omega-\omega')\\
&\int dt(\bar v_\omega(t)u_{\omega'}(t)+\bar
u_{\omega}(t)v_{\omega'}(t))=0
\end{align}

Since by assumption the dispersive terms are
dynamically dominant, the eigenfunctions of $L$ are close to the
eigenfunctions of the linearized NLS equation \cite{kaup}
\begin{align}
u_\omega(t)&=e^{i\omega t}\left(1-\frac{2i\omega e^{-a t}}{(\omega+ia)^2}a\sech(at)
+\frac{a^2\sech^2(at)}{(\omega+ia)^2}\right)\label{eq:uomega}\\
v_\omega(t)&=e^{i\omega t}\frac{a^2\sech^2(at)}{(\omega+ia)^2}\\
\bar u_\omega(t)&=a\frac{(\omega+ia)^2}{(\omega-ia)^2}u_{-\omega}(-t)\\
\bar v_\omega(t)&=-a\frac{(\omega+ia)^2}{(\omega-ia)^2}v_{-\omega}(-t)
\end{align}
so that
\begin{align}
\beta_{g\omega}&=\frac{a}{2\pi}\frac{(\omega+ia)^2}{(\omega-ia)^2}\int
dt(1+\gamma a^2\partial_\theta^2)a\sech\theta
\\\nonumber&\quad\times e^{-i\omega t}\Big(1-\frac{2i\omega e^{-a
t}a\sech(at)+{a^2\sech^2(at)}}{(\omega+ia)^2}
\Big)\\
\beta_{s\omega}&=\frac{a}{2\pi}\frac{(\omega+ia)^2}{(\omega-ia)^2}\int
dt\big(s(a^2\sech^2(at))-l\big)a\sech(at)
 \\\nonumber&\quad\times e^{-i\omega t}\Big(1-\frac{2i\omega e^{-a t}a\sech(at)+{a^2\sech^2(at)}}{(\omega+ia)^2}
\Big)
\end{align}

\smallskip

The solution of Eq.\ (\ref{eq:dza}) is
\begin{align}\label{eq:chi}
\alpha_\omega(z)&=\int^zdz'e^{(-\omega^2(i+g\gamma)-ia^2)(z-z')}\\\nonumber&\times
e^{\int_{z'}^zdz''(gp_g(z'')-lp_s(z''))}(\beta_{g\omega}
p_g(z)+\beta_{s\omega}p_s(z))
\end{align}
As observed by Gordon and Kelly \cite{Gordon,kelly1,kelly2}, the
amplitudes $\alpha_\omega$ are driven resonantly if the nonlinear
frequency shift $a^2+\omega^2$ is an integer multiple of the cavity
based wavelength $\frac{2\pi}{L}$; spectrally, therefore, the
dispersive waves contain a discrete set of sidebands at frequencies
$\pm\omega_n$,
\begin{equation}
\omega_n=\sqrt{\frac{2\pi}{L}n-a^2}\ ,\ \  n=1,2,\ldots
\end{equation}
The $n$th sideband is forced mainly by the $n$th harmonic of the
gain and loss modulation functions $p_{g,l}$, but also by the $z$
dependence of the of the gain and loss terms in the exponent in Eq.\
(\ref{eq:chi}). The latter modulation is small, however, if we make
the simplifying assumption that the total net loss per roundtrip is
small, so that the $z'$ integration in Eq.\ (\ref{eq:chi}) is
effective over many roundtrips. This is the the usual assumption
underlying the Haus master equation \cite{HausReview}, and is also
consistent with the relative weakness gain and loss processes in the
dynamics dominated by the chromatic dispersion and Kerr nonlinearity
that is studied here.

In this case the
mean net loss in the $z''$ integration in Eq.\ (\ref{eq:chi}) dominates over the variable parts of the gain and loss, and the latter may be neglected. The same assumption justifies the approximation of the time-dependent linear operator in Eq.\ (\ref{eq:pert1}), the equation of motion for $\psi_b$, by the fixed operator $L$ that was used to derive Eq.\ (\ref{eq:dza}). 

It now follows that the dispersive wave amplitudes are
\begin{align}\label{eq:chi-sb}
\alpha_\omega(z)&=\int^zdz'e^{(-i (a^2+\omega^2)+g(1-\gamma \omega^2)-l)(z-z')}\nonumber\\
&\quad\times(\beta_{g\omega} p_g(z)+\beta_{s\omega}p_s(z))
\end{align}
and defining the $n$th Fourier components $\tilde p_{g,n}$ and
$\tilde p_{s,n}$ of $p_g$ and $p_s$ (respectively), the coefficients
of the $n$th sideband in the steady state are
\begin{equation}\label{eq:chin}
\alpha_{\omega,n}(z)=\frac{(\tilde p_{g,n}\beta_{g\omega}+ \tilde
p_{s,n} \beta_{s\omega})e^{-\frac{2\pi}{L}inz}}{l+g(\gamma
\omega^2-1)+i(\omega^2 -\omega_n^2)}
\end{equation}

We wish to characterize the sidebands by their ordinary Fourier
spectrum, obtainable from Eq.\ (\ref{eq:alpha}),
\begin{equation}\label{eq:psin}
\tilde\psi_{bn}(\omega,z)=e^{ia^2z}\int\frac{d\omega'}{2\pi}(\alpha_{n\omega'}(z)\tilde
u_{\omega'\omega}+\alpha_{n,\omega'}(z)^*\tilde v_{\omega'\omega}^*)
\end{equation}
where
$\tilde u_{\omega'\omega}=\int dte^{-i\omega t}u_{\omega'}(t)$, and $\tilde
v_{\omega'\omega}$ is defined similarly.
As a transform of a rapidly varying function, $\tilde u$ is
wideband---its $\omega$ bandwidth for fixed $\omega'$ is comparable
with the soliton bandwidth---but it is also singular for
$\omega=\omega'$.  The smooth part of $\tilde u$ generates a small
deformation of the soliton waveform in Eq.\ (\ref{eq:psin}) that is
unimportant for the present purpose of characterizing the sideband
spectrum. We therefore focus on the singular part $\check
u_{\omega'\omega}$ of $\tilde u_{\omega'\omega}$ that is determined by the
$\pm\infty$ asymptotes of $u_{\omega'}$ (see Eq.\ (\ref{eq:uomega}) ) to
\begin{equation}
\check u_{\omega'\omega}=2\pi\frac{\omega^2-1}{(\omega+ia)^2}\delta(\omega-\omega')+
\frac{4\omega a}{(\omega+ia)^2}\mathsf{P}\frac{1}{\omega-\omega'}
\end{equation}
where $\mathsf{P}$ denotes principal part. $\check
v_{\omega'\omega}=0$ since $v_{\omega}(t)$ tends to zero
exponentially for $t\to\pm\infty$.

Now we can carry out the $\omega'$ integration in Eq.\
(\ref{eq:psin}) and obtain the spectrum of the $n$th sideband
\begin{equation}\label{eq:final_c}
\tilde\psi_{b,n}(\omega)=\frac{(\tilde p_{g,n}\beta_{g\omega}+\tilde p_{s,n} \beta_{s\omega})e^{i(a^2-\frac{2\pi}{L}n)z}}{l+g(\gamma \omega^2-1)+i(\omega^2 -\omega_n^2)}
\end{equation}

The spectral width (half width at half maximum) of the sideband is
given by $\delta\omega=(l+g(\gamma\omega_n^2-1))/2\omega_n\ll1$;
temporally, therefore, the dispersive wave is a pedestal of
frequency  $\omega_n$ with an exponentially decaying envelope
centered at the pulse, whose decay time scale is much wider than the
soliton width.

\begin{figure}[htb]
\hbox{\includegraphics[width=9cm]{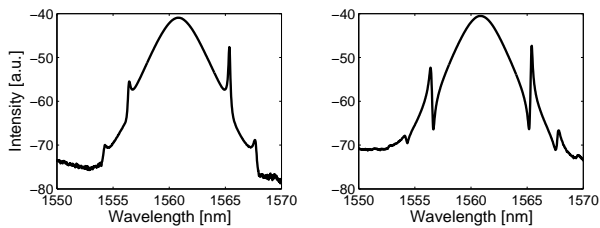}} \caption{Measured
spectra of the soliton and its sidebands, showing the notch
formation. The two figures correspond to two measurement locations
in the same cavity that are separated by one quarter of cavity
length. These mesurements agree well with the
theory.\label{fig:spec}}
\end{figure}

\begin{figure}[htb]
\hbox{\includegraphics[width=8cm]{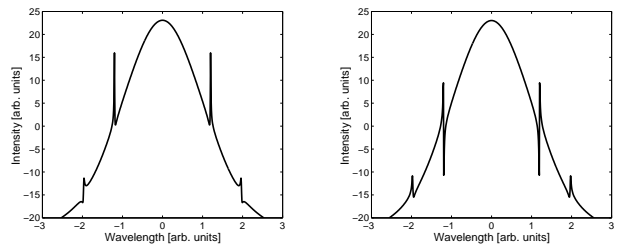}}
\caption{Theoretical calculation of the pulse and sideband spectrum with: $a=1, g=0.1, \gamma=0.01,
l=0.11, s=0.1(|\psi|^2-|\psi^4|), t_R T=10^{-3}, L=2.56,
\tilde p_g=e^{i\pi/2}, \tilde p_s=e^{i\pi}$. The cavity position in the left
figure is $z=L/5$, and in the right is $z=0.
$\label{fig:spec_theory}}
\end{figure}

Experimentally, the sidebands appear as a series of sharp peaks
in the pulse spectrum on the background of the wide soliton
spectrum, (see Fig.\ \ref{fig:spec}). A distinctive feature of the
sideband spectrum (\ref{eq:final_c}) is the dependence of the
overall phase of the sideband on the measurement position in the
cavity; as a consequence, the sideband spectrum interferes
constructively or destructively with the soliton spectrum, depending
on the placement of the output coupler, exhibiting a sharp notch
feature for some placements as the result of destructive
interference. Figs.\ \ref{fig:spec}--\ref{fig:spec_theory} show the
comparison between the measurement and calculation of the soliton
and the first two sidebands spectrum for two cavity positions. The
main qualitative discrepancy between theory and experiment is the
asymmetry between the left and right sidebands that is lacking in
the theory. The most likely source of this discrepancy is the
assumption of a nonchirped zero frequency pulse, that only
approximately holds in the experiments.

\paragraph{The continuum component}
The continuum $\psi_c$ is the dominant component of the waveform for
most of its temporal extent, where both $\psi_p(t)$ and $\psi_b(t)$ are
negligible. Therefore, the nonlinearity and interaction with the
pulses are unimportant for its dynamics, and it is natural to
express it in terms of the ordinary Fourier modes
$\tilde\psi_c(\omega)$ that satisfy
\begin{equation}\label{eq:psi_n}
\partial_z\tilde\psi_c=-i\omega^2\tilde\psi_c
+p_g(z)g\left(1-\gamma\omega^2\right)\tilde\psi_c
-p_s(z)l\tilde\psi_c+\tilde\Gamma(\omega,z)
\end{equation}

Because the noise and continuum extend throughout the entire
roundtrip time, we must use discrete frequencies to label their Fourier
transform. It then follows from Eq.\ (\ref{eq:gamma}) that
$\tilde\Gamma(\omega,z)==\int dte^{-i\omega t}\Gamma(t,z)$ has zero mean and correlation function
$\langle\tilde\Gamma(\omega,z)\tilde\Gamma^*(\omega',z')\rangle=2T
t_R\delta(z-z')\delta_{\omega,\omega'}$. The solution of Eq.\
(\ref{eq:psi_n}) is similar to (\ref{eq:chi}), and by the same
arguments that lead to Eq.\ (\ref{eq:chi-sb}) we again approximate
the integrand in the exponent by its mean value. The resulting
expression implies that $\langle\tilde\psi_c(\omega)\rangle=0$ and
\begin{equation}\label{eq:psi_n2}
\langle\tilde\psi_c(\omega)\tilde\psi_c^*(\omega')\rangle=\frac{t_R
T\delta_{\omega,\omega'}}{l+g(\gamma\omega^2-1)}
\end{equation}

\paragraph{Gain balance and multipulse sidebands}
In the analysis presented so far, the pulse parameters and the overall
saturated gain $g$ were assumed fixed and given. In the steady state
these are variables, determined along with the number of pulses as a
solution of the optical equation of motion Eq.\ (\ref{eq:master}).
Since the equations of motion are random, the result is a
\emph{statistical} steady state. In previous works \cite{GGFPRE,
PRL-2006}, statistical light-mode dynamics (SLD) theory was used to
study this problem, and applied to multipulse mode locking in
\cite{stepsPRL,stepsPRE}. The global mode locking analysis
determines in particular the properties of the sidebands, and in
this way allows us to reach our goal of describing the width and
power of the sidebands as a function of the laser parameters.

Here we study the statistical steady state by the gain balance
method, deriving equations of motion for the power in the three
components of the optical waveform, that is, the power $P_p$ in the
pulse waveform that comprises zero or more pulses, the sideband
power $P_b$, and the continuum power $P_c$. The three waveform
components are characterized by well-separated time scales, and the
total power can therefore be calculated as a sum of the powers of
the individual components.  One looks for a steady state where the
three components are subject to the same saturated gain, and finally
the gain itself is determined self-consistently from Eq.\
(\ref{eq:gain}).

We will assume that the multipulse waveform is of the simplest kind
\cite{stepsPRL} that is also the most commonly observed
experimentally, consisting of $k$ pulses of equal amplitude $a$. In
the equation of motion for the pulse amplitude we may neglect the
noise term in the master equation, and use the results of soliton
perturbation theory \cite{Gordon,karpman} to write
\begin{align}
\frac{da}{dz}&=\re\int dt \psi_p^*\bigl(gp_g(z)(1+\gamma_g\partial_t^2)
\psi_p\nonumber\\&\qquad+ p_s(z)(s(|\psi_p|^2)-l)\psi_p\bigr)\nonumber\\
&=2gp_g(z)a(1-\frac{\gamma}{3}a^2)+p_s(z)(2s_a(a)-2l)a\label{eq:dadz}
\end{align}
where
\begin{equation}\label{eq:nonlinear_gain1}
{s}_a(a)=\frac{1}{2a}\int
s(a^2\sech^2(at))a^2\sech^2(at) dt
\end{equation}
The pulse amplitude changes periodically during its propagation in
the cavity; for our purposes we need the mean pulse amplitude $\bar
a(z)=\frac1L\int a(z+z')dz'$. Under the assumption of weak gain and
loss processes the dynamics of $\bar a$ is obtained by the space
average of Eq.\ (\ref{eq:dadz}), that gives in the steady state
\begin{equation}
g(1-\frac{\gamma}{3}\bar a^2)-l+s_a(\bar a)=0
\end{equation}
This equation, together with $P_p=2k\bar a$, determines the pulse
part of the gain balance.
\begin{figure}[tb]
\hbox{\includegraphics[width=9cm]{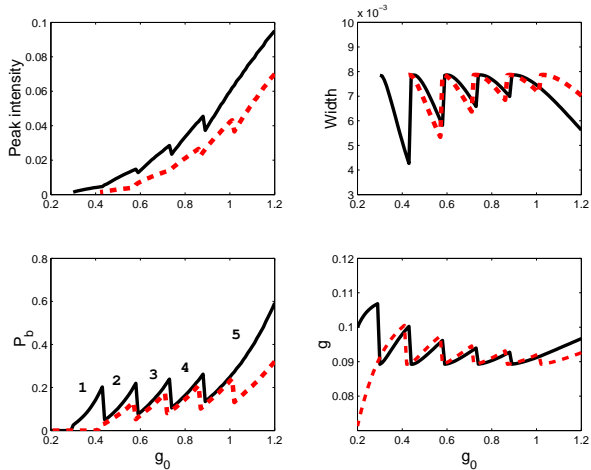}}
\caption{The theoretical calculated values of the sidebands peak
intensity, width, and total power $P_b$ along with the saturated
gain coefficient $g$ as a function of the small signal gain $g_0$.
The same parameters as in Fig.~\ref{fig:spec_theory} were used in
the calculation, except that $\gamma=10^{-3}$ and $P_S=1$. Two noise
injection rate values are shown: $T=10^{-3}$ in continuous black
line, and $T=3\times 10^{-3}$ in dashed red line. The number of
pulses is indicated.
All values are given in natural soliton units. \label{fig:theory}}
\end{figure}

Next, we calculate the power carried by the pedestals of the pulses.
Each pulse generates a series of sidebands of the form given by Eq.\
(\ref{eq:final_c}). The sidebands power is dominated by the leading,
$n=\pm1$ sidebands; since the different pulses that act as sources
for the sidebands are not phase locked, we treat them as incoherent
sources, and accordingly calculate the total sidebands power, as the
sum of the individual sideband powers. The resulting expression is
\begin{equation}\label{eq:side_p}
P_{b}= k\frac{|\tilde p_{g,1}\beta_{g\omega_1}+\tilde
p_{s,1}\beta_{s\omega_1}|^2}{2\omega_1(l+g(\gamma\omega_1^2-1))}
\end{equation}

\begin{figure}[tb]
\hbox{\includegraphics[width=9cm]{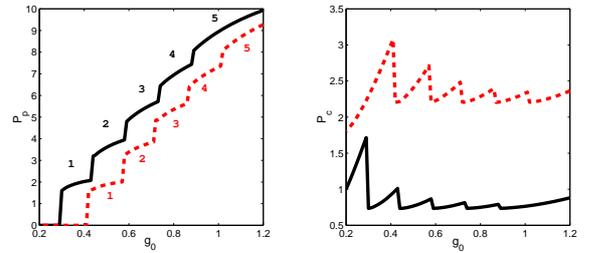}} \caption{The
theoretical calculated values of the total pulse power $P_p$ and noise power $P_n$ as a function of the small signal gain $g_0$.
The same parameters as in Fig.~\ref{fig:theory}. The number of pulses is indicated. All values are given in
natural soliton units.\label{fig:pulse_and_noise}}
\end{figure}

Finally the mean power of the the continuum is not changed by the
presence of the sidebands, and is therefore given  by the expression
derived in \cite{njp}
\begin{equation}\label{eq:noise_p}
P_c=\frac{Tt_R}{2\sqrt{g\gamma (l-g)}}
\end{equation}
smaller than the pulse power $2a$, so that $P_c$ is of the same
order of magnitude as $P_s$.

We now substitute $P=P_s+P_{b}+P_c$ into Eq.~(\ref{eq:gain}) and
obtain a nonlinear equation for the saturated gain $g$ that is easy
to solve numerically. Once we know the value of $g$ we obtain the
steady state solution of the entire waveform. The steady state
equation can have several solutions with different numbers of pulses
$k$. In such cases the laser waveform can exist in several states
\cite{stepsPRL,stepsPRE}, in a similar manner to the existence of
metastable phases in thermodynamic systems undergoing first order
phase transitions. The laser then exhibits hysteresis---its actual
state depends on the history.

Figs.~\ref{fig:theory}--\ref{fig:pulse_and_noise} show the results
of the calculation as the small signal gain $g_0$ is varied. For
very low gain the waveform is pure continuum, and its energy
increases with the gain. When $g_0$ is increased beyond a certain
threshold, a pulse forms and along with it also a dispersive wave.
Because of gain saturation, the net gain must abruptly decrease, and
along with it also the continuum component. Further increase of
$g_0$ will mainly increase the sidebands and the continuum
components, and slightly change the pulse amplitude, until the
second threshold is met, and then the continuum as well as the
sidebands power abruptly decrease again. This process then continues
periodically when further pulse creation thresholds are reached. In
addition, as the pumping is increased between pulse creation
thresholds, the saturated gain increases, so that the net loss
decreases and the sidebands become spectrally narrower, and
accordingly temporally wider.

\begin{figure}[tb]
\hbox{\includegraphics[width=9cm]{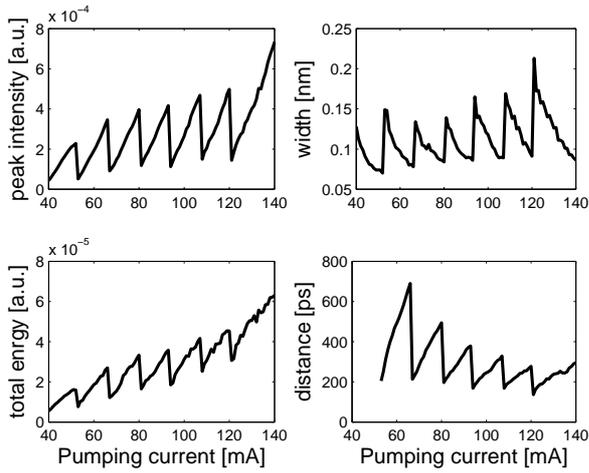}}
\caption{The measured sideband spectrum characteristics: peak power, spectral
width, and total energy, and the distance between
adjacent pulses in states with two or more pulses.}\label{fig:exp}
\end{figure}

The theoretical predictions of the global sidebands characteristics
agree well with experimental observation summarized in
Fig.~\ref{fig:exp}, made on a fiber ring laser  mode locked by
polarization rotation as described in \cite{stepsPRE}. In both
graphs the sideband width displays roughly periodic sawtooth
behavior and the total sidebands energy a sawtooth behavior with an
increasing linear bias. There is somewhat worse agreement in the
peak intensity, but this quantity is sensitive to interferometric
enhancement and reduction as explained above.

In both theory and experiments the spectral width of
the sidebands narrows with increasing pumping between pulse creation
thresholds, implying a temporal widening of the pedestals. This widening shows a striking
correlation with the increase of the inter-pulse distance,
suggesting that the sidebands play a role in the inter-pulse
interactions \cite{Grudinin_Gray, Soto-Crespo}.

\paragraph{Conclusions and outlook}
We presented a fundamental theory of the formation of sidebands and
their effect on the statistical steady state in multipulse
mode-locked soliton lasers, that explains the main experimental
observations, including the dependence of the sideband spectrum on
the measurement position in the cavity, the growth of the sideband
energy and temporal width when pumping is increased, and the abrupt
attenuation of the sidebands when a new pulse is created in the
cavity. We also found strong correlations between the temporal width
of the sidebands and the spacing between adjacent pulses in pulse
bunches, giving further evidence for the role of the sidebands in
inter-pulse interaction. However, the most natural conclusion from
our observation is that the sidebands generate \emph{repulsive}
interactions, and that additional attractive interactions are needed
to explain the ubiquitous formation of pulse bunches. Moreover, the
phase and timing jitter of the pulses leads us to conjecture that
the mechanism of interaction is {incoherent}. We postpone the
in-depth study of these question to a future publication.

Acknowledgments: This research was supported by the Israel Science
Foundation.

\end{document}